# Population-scale testing can suppress the spread of infectious disease

Jussi Taipale*†+✉, Ioannis Kontoyiannis+, and Sten Linnarsson†✉

*† Department of Medical Biochemistry and Biophysics, Karolinska Institutet, Stockholm, Sweden.*
*\*Applied Tumor Genomics Research Program, Faculty of Medicine, University of Helsinki, Finland and Department of Biochemistry, +University of Cambridge, Cambridge, United Kingdom.*

✉e-mail: sten.linnarsson@ki.se; ajt208@cam.ac.uk

**Major advances in public health have resulted from prevention of disease. However, prevention of a new infectious disease by vaccination or pharmaceuticals is made difficult by the slow process of vaccine and drug development. We propose here an additional intervention that would allow rapid control of emerging infectious diseases, and could also be used to eradicate diseases that rely almost exclusively on human-to-human transmission. The intervention is based on: (1) testing every individual for the disease, (2) repeatedly, and (3) isolation of infected individuals. We show here that at a sufficient rate of testing and isolation, the reproduction number $R_0$ is reduced below 1.0, and the epidemic will rapidly collapse. The approach does not rely on strong or unrealistic assumptions about test accuracy, compliance to isolation, population structure or epidemiological parameters, and its success can be monitored in real time by measuring the change of the test positivity rate over time. In addition to the rate of compliance and false negatives, the required rate of testing is dependent on the design of the testing regime, with concurrent testing outperforming random sampling of individuals. Provided that results are obtained rapidly, the test frequency required to suppress an epidemic is monotonic and near-linear with respect to $R_0$, to the infectious period, and to the fraction of susceptible individuals. Importantly, the testing regime would be effective against both early phase and established epidemics, and additive to other interventions such as contact tracing and social distancing. We show that the approach is also robust to failure – any rate of testing reduces the number of infected individuals in the population, improving both public health and economic conditions. These conclusions are based on rigorous analysis as well as simulations of appropriate epidemiological models. A mass-produced, disposable antigen or genetic test that could be used at home would be ideal, due to the optimal performance of concurrent tests that return immediate results.**

Very broadly speaking, an outbreak of a novel infectious disease in a naïve population can have only two outcomes, early extinction, or widespread epidemic affecting a large fraction of the population. Initially, when one or very few individuals are infected, even an epidemic of a highly infectious pathogen (virus hereafter) can die out because of random fluctuations. The most important determinant for the probability that an epidemic outgrows this early phase is the probability that an infected individual infects at least one other individual ($1-p_0$). After the infection takes hold, the course of the infection is determined by another summary statistic, the reproduction number $R_0$ of the virus, which represents the expected value of the number of secondary infections caused by an infected individual. As long as the reproduction number $R_0$ remains greater than 1.0, the virus spreads rapidly until most people have been infected (**Fig. 1A**), creating a temporary surge of infected individuals.

If, using pharmaceutical or social interventions, $R_0$ can be reduced below 1, then the epidemic collapses (**Fig. 1B**), and most people remain uninfected (but still susceptible). Because of the exponential nature of epidemics, the outcomes are nearly binary. Even when $R_0$ exceeds 1.0 by only a small amount the disease spreads at an accelerating pace, whereas as soon as $R_0$ falls just below 1.0 it rapidly collapses. These two outcomes correspond to two distinct strategies for epidemic control, suppression and mitigation.

In the mitigation model, the goal is to reduce $R_0$ as much as possible but not necessarily below 1.0, hoping to end up with a population that is largely immune, without overwhelming the healthcare system in the process (as in **Fig. 1A**, but attempting to flatten the temporary surge of infected individuals). If the virus induces immunity, the mitigation approach also limits the total number of people infected, and leads to "herd immunity" (see, for example Refs. [1,2]), which would limit future epidemics caused by variants of the same virus. However, exponential processes are notoriously difficult to control, particularly in the absence of accurate real-time data and when the effect of policy changes is uncertain. The choice is stark: allowing the disease to spread to a large fraction of a population, however slowly, greatly increases the total number of infected people and would cause a severe loss of life. Furthermore, given the difficulties in controlling exponential processes using limited information, even a strongly enforced mitigation strategy runs the risk of overwhelming the health care system and significantly increasing the mortality rate due to the failure to treat every patient optimally (primarily due to the lack of intensive care units' capacity). If the healthcare system is overwhelmed, patients must be triaged as in wartime, potentially for extended periods of time.

At low levels of prevalence, testing, contact tracing and quarantine (TTQ) is a very effective means of suppression (Refs. [2,3]), because it reduces the effective rate of reproduction close to zero. It is not an efficient strategy for suppressing an epidemic with significant community transmission, as it requires extensive resources that scale linearly with the number of infected individuals. In addition, the TTQ approach depends on a population-scale identification of a subset of all infected individuals – currently this is largely based on testing of symptomatic individuals. Therefore, the approach is not effective in suppressing infectious agents that can be transmitted by asymptomatic individuals. This is because asymptomatic individuals that are not identified will generate new clusters that will not be detected until someone develops a severe infection that requires medical care. As a result, suppressing an epidemic using the TTQ approach takes a long time and is also potentially unstable. Once the rate of new cases exceeds the capacity of tracing, even briefly, the epidemic grows out of control and the exponential dynamics make it almost impossible to catch up with without imposing a lock-down.

In summary, current approaches to epidemic control rely on fast action by indiscriminate lockdown or TTQ, and on the slower process of development of vaccines and pharmaceutical treatments. Given the limitations of these approaches, we asked if there are methods to combat emerging infectious diseases that would be less costly than a lockdown, and that could be developed faster than vaccines or effective therapies.

Here, we evaluate a strategy that is both simple and fast: testing everyone, repeatedly. When someone tests positive, ask them to self-isolate and provide them public assistance that reduces the burden this imposes on them. This approach relies on a key observation that has not been widely appreciated, namely that what matters is the fraction of all individuals that are identified and isolated. It follows that testing a small number of individuals with a highly accurate test and isolating them with high compliance can be much less effective than testing everyone with a less accurate test, and isolating them with lower compliance. In fact, there is a quantifiable relationship between the reproduction number of a virus, and the efficiency of a population-scale testing strategy that brings the effective reproduction number below 1.0, as we show below. We use deterministic as well as stochastic epidemiological models to derive explicit expressions as well as upper and lower bounds on the critical rate of testing to suppress both early and established epidemics, and the marginal utility of testing at lower rates.

This approach has several important advantages. First, it will work no matter how high the prevalence of infection might be. Second, it does not suffer from the inherent instability of contact tracing that requires a very high speed and a secondary mechanism to catch lost contacts (e.g. testing symptomatic individuals). The offsetting disadvantage is that it is a challenge to test at the required scale, but this is not as difficult as it might at first seem. It could be implemented using mass distribution of self-administered tests via regular mail, without returning samples to a central testing site. In fact, the tests required do not even have to be properly "diagnostic", as in the simplest case the test result would only influence the decision to self-isolate. In the worst case, this may cause people who are not infectious to be quarantined, but this is already true for most people in a lockdown scenario. This regime can tolerate false positives, because the result of a provisionally positive test is that someone self-quarantines for a relatively short period of time when they did not have to. False negatives are also acceptable as long as people are retested frequently.

The proposed approach is empirical, and does not depend on complex epidemiological models or a highly time-effective, centrally directed response – although, as we show, its efficacy is indeed supported by epidemiological models. It only requires knowledge of the nucleic acid sequence of the threat or some other antigen, e.g. a protein, that is relatively abundant in the pathogen, and can thus be implemented with very limited initial understanding of an emerging new infectious disease. Provided that relevant information is collected, the process dynamically and efficiently generates the evidence required for optimization of parameters and countermeasures. The simplicity of the process and the effective accumulation of evidence are very important considerations in epidemics, where rapid action to limit spread is critical, and where minimizing the number of cases – the source of the primary evidence – is a central objective. The success of the programme can be assessed simply by evaluating the test positivity rate over time, and efficacy can be optimized at an almost arbitrary level of granularity based on positivity rates as a function of other parameters.

To guide initial design, we used the deterministic susceptible-infected-recovered (SIR) model to examine the effects of false negatives and noncompliance (see **Methods**). We first make a best-case assumption about the timing of the tests: every person who is infected is tested before encountering someone who is susceptible. This limit can be approached, for example, when testing individuals as a condition of release from quarantine, or by a very effective form of contact-tracing. Note that if a perfectly accurate test were applied to the entire population at once, and those who tested positive were fully isolated, the epidemic would immediately collapse with no new infections (**Fig. 1C**). This optimal population-scale testing strategy will succeed in collapsing the epidemic if the fraction of infected persons who are isolated exceeds $(R_0 - 1)/(R_0 - R_q)$, where $R_q$ is the reproduction number in isolation or quarantine (**Fig. 2A** and **Methods**). For example, if $R_0 = 2.4$, $R_q = 0.3$, $p$ is the fraction of true positives correctly identified by the test, and $c$ is the fraction of the public that complies in the sense that they agree to be tested and follow any instruction to go into isolation, this bound means that the product $cp$ must be greater than 2/3.

Next, we assume instead that the test sensitivity and compliance are perfect, so $p = c = 1$, and consider a reasonable worst-case assumption on the timing of the tests: each day, a randomly selected fraction of the population is tested. Under that strategy, we find that testing at a rate greater than a fraction $(R_0 S/N - 1)$ of the population per infectious period will ensure that the effective reproduction number $R_e < 1$ (**Fig. 2B**, **Methods**; S/N is the fraction still susceptible). Real-world testing strategies could do much better than test at random, for example by implementing procedures that test individuals concurrently within a region; that run the screen as a sweep across a country; that slice the population into groups that are tested in a cycle; that test individuals that have many contacts, or use other variables to predict who is more likely to be infected or more likely to infect others and to test them more frequently (see **Methods**). The theoretical limit of performance of most testing strategies lies between the two bounds delineated above. The order of performance is defined by a simple order rule (see **Methods**), and is: testing the vector (e.g. blood before transfusion), testing before infectiousness, testing all people at the same time, testing each person but at different times, and random sampling of individuals. Cyclic processes that occur naturally (e.g. activity cycle across a week) or due to policy (e.g. rolling lockdown[4]) will also partially synchronize infectious intervals, and facilitate the application of the stronger testing regimes. Furthermore, if local or population-scale information exists that can be applied for contact-tracing, the efficacy of the regime can be significantly improved by isolating or quarantining the contacts of the individuals who test positive.

The fact that the approach will work is clear if one considers, e.g. in the case of Covid-19, that the current approaches are "natural" variants of a population-scale testing regime. For example, lockdown corresponds to a test with a sensitivity of zero and false positive rate of one, followed by a quarantine that is applied at a relatively low "compliance" (for example, many essential workers are often exempt). Similarly, isolation of symptomatic individuals corresponds to a non-biochemical test that measures presence of infection based on self-assessment of generic symptoms, leading to a relatively low sensitivity and specificity, and a suboptimal timing (for example, a self-assessed or even expert-guided symptom-test for SARS-CoV-2 is centered at the middle of the infectious interval, and will not detect cases that remain non-symptomatic throughout the infectious period). These parameters can clearly be improved by an introduction of a population-scale biochemical test. Precisely by how much depends on the baseline approach, and parameters such as fraction of successfully traced contacts in the population, timing of symptoms relative to infectiousness, and the effect of test results on compliance (see **Methods**). Furthermore, the rate of testing required to suppress an epidemic is also sensitive to other parameters. However, despite the exponential nature of the epidemic itself, the sensitivity of the required testing interval to variation in most parameters is linear or nearly so. For example, both a shorter infectious period and a larger $R_0$ shorten the require testing interval proportionately. The most impactful parameter is the delay $d$ in obtaining a test result, which should be minimized in any realistic plan for population-scale testing. This is because if $d$ is longer than the sum of the detection lead time ($l$) and $R_0$ divided by the infectious period, the epidemic cannot be suppressed with any rate of testing in the absence of other interventions (see **Methods**).

Population-scale testing positively interacts with other strategies. Interventions that reduce $R_0$ — e.g. working from home, improved hand hygiene, isolation even with mild symtoms, the use of masks and social distancing — are additive with respect to the testing, and hence lower the required frequency of the tests. Similarly, as the epidemic progresses and fewer people remain susceptible, the frequency of testing required to control the epidemic drops. The relationship between these variables in the SIR model with testing is captured by the following inequality (see **Methods**), which relates the testing rate ($\tau$, average fraction of population tested per day), the recovery rate ($\gamma$, reciprocal of the infectious period duration), and the proportion still susceptible (S/N): $\tau > \gamma(R_0 S/N - 1)$. As shown in **Fig. 3**, with $\gamma = 1/5$, assuming moderate non-pharmaceutical interventions that reduce $R_0$ to the range $1.2 - 1.5$, and assuming most (90%) of the population is still susceptible, population-scale testing on average every 15 – 120 days would be sufficient to bring the effective reproduction number below 1.0 and thus control the epidemic. With much longer (14 days) or shorter (3 days) infectious intervals, the required rate of testing changes accordingly. Importantly, in a realistic range of parameters, it is particularly difficult to find values that would both make it impossible to suppress the epidemic by population-scale testing, and prevent relatively easily implementable alternative means of suppression (wherein testing would commonly also be an important component). For example, if the infectious period was only two days long, suppression of the epidemic by random testing alone would be difficult – but instead a short quarantine followed by a test would be highly effective. Similarly, if individuals with short and long infectious intervals co-exist in the population due to genetic or environmental variation, the short duration infections can be suppressed by a lockdown, and the long ones by testing. The figure also illustrates the trade-offs involved: if population-scale testing rates can be increased from once every four months (1/120$^{th}$ of the population per day) to once per two weeks (1/14$^{th}$ of the population per day), then alternative measures such as lockdowns can be reduced from $R_0 \approx 1.2$ to $R_0 \approx 1.5$.

The standard but simple and deterministic Susceptible-Infectious-Recovered (SIR) model used to calculate these bounds is based on strong assumptions and approximations, such as random mixing of all individuals, perfectly accurate testing, full compliance and perfect isolation of cases.

To relax those assumptions, we first implemented a more realistic numerical simulation using a stochastic model with testing. The deterministic model above predicted that testing at least every ten days would be required to control an epidemic when $R_0 = 1.5$. **Fig. 4** shows a simulation that starts with one infected individual in a population of 10,000. The analysis confirms that concurrent testing outperforms a random approach, and that both approaches are significantly improved by quarantining contacts of the infected individuals. The random approach can also be improved by targeting the tests to high-contact individuals. The simulation also confirms that delays in reporting test results lead to significant degradation of performance of the approach. Even a short delay in reporting of test results (48 hour delay when the infectious period is 7 days) degrades performance as much as a decrease in true positive rate of the test from 1 to 0.5.

It is important to note that the benefits of the testing regime are very counterintuitive in that they scale with the size of the total human population – not with a much more intuitive number, the number of infected individuals within the population (which can be very small). This is clear if one, for example, considers a scenario where a population-scale testing regime had caught the first individual infected with SARS-CoV-2, for example. It would have immediately collapsed the epidemic, saved hundreds of thousands of lives, and prevented the world economy from entering a severe downturn whose costs far exceed the cost of testing everyone on the planet with ~ 7.8 billion biochemical tests.

To further relax our assumptions, we consider a general, randomized Reed-Frost epidemic on a random network modelled either as a Bernoulli random graph or as a random graph with a given degree distribution (known as the "configuration model"); see **Supplementary Methods**. Rigorous analysis gives generalizations of the previous findings described, and allows consideration of further issues.

In particular, equally importantly to considering the probability of success, a medical decision affecting public health must also weigh the consequences of failure. In the general case where testing is insufficient and $R_0$ remains above 1.0 despite all interventions, the benefit from the testing would be the difference between the total size of the epidemic at original $R_0$ and the effective $R_{\text{eff}}$ under the testing regime. Mathematical analysis reveals that any rate of testing leads to a decrease in number of infections (**Fig. 5**; for detailed analysis, see **Supplementary Methods**). At an early stage of an epidemic, the probability of extinction of the virus is increased linearly as a function of testing rate, reaching 1 at a critical rate that is explicitly identified as a function of the underlying epidemic parameters (**Fig. 5A**, **C**). Similarly, any rate of testing leads to a decrease in the "attack rate" – the fraction of people infected at the end of the epidemic (**Fig. 5B**, **D**). We conclude that, under very general assumptios, the consequence of failure to fully implement the population-scale testing and isolation program would still lead to improvement in both public health and economy, a finding that does not depend on the fractional scale at which the regime is implemented.

There are many plausible technical approaches to population-scale testing. Such a test can be based on presence of virus antigen, either in the form of viral proteins[5,6], or viral RNA like current state-of-the-art diagnostic tests (for example, Ref. [7]). In both cases, a home self-test is preferable due to the simple logistics and quick time to result, which reduces the crucial latency to isolation of infectious individuals. Despite the technical simplicity, it is difficult to translate current tests designed for medical diagnostic purposes to a field setting. Current diagnostic tests for pathogens such as SARS-CoV-2 are quantitative reverse-transcription polymerase chain reaction (qRT-PCR) assays that require (1) nasopharyngeal swab collected by a trained nurse, (2) sample collection in viral transport media, (3) RNA purification, (4) reverse transcription and quantitative PCR. The test is highly accurate, and the total cost is in the order of $100 US. Such highly accurate testing is critical for accurate diagnosis of infections in a hospital setting. However, due to the very detailed and specific regulation, specialized staff and equipment, and centralized testing facilities, such tests have proven difficult to rapidly scale above thousands of assays in each location. A distributed system of sample collection and testing can, however, be used to scale qRT-PCR and antigen-tests to population levels, as recently demonstrated in China and Slovakia, respectively (Ref. [6]). The capacity can also be increased 10- to 100-fold by group testing[8], a method with a long history of use in public health that was originally designed for Syphilis tests, and now commonly also used for optimally efficient detection of defective components in industrial production.

A parallel relatively centralized testing method based on existing DNA sequencing technology could also be fielded rapidly. In this approach, viral RNA in the samples is used to generate DNA sequences containing the virus sequences, a sample DNA barcode (to identify each case) and two unique molecular identifiers[9] at both ends of the resulting DNA fragment (to count the number of virus RNAs per sample and to ensure that patient samples do not get mixed in the reaction), and then sequenced using a massively parallel sequencer. This approach is very scalable as, in principle, a single sequencing instrument that is routinely used in scientific research can report more than a billion results per day. Furthermore, in the future, a test based on sequencing (Refs. [10-12]) that covers many acute infections could also be used to suppress or even eradicate a large number of infectious diseases simultaneously. This would be very difficult to achieve using vaccines or drugs that target each infectious agent separately.

Alternatively, we envisage supplementing the current testing regime with a mass-produced home test kit that could be used by anyone, result in a simple easily-understood readout, and be performed without specialized equipment. The test should be as easy to use as a pregnancy test to ensure maximal compliance; importantly, using a home test, the time delay to report the result would be effectively eliminated. Boxes of e.g. 50 tests would be mass-mailed to all citizens, and a national information campaign would encourage everyone to test themselves frequently. In an infected individual, viral RNA is present at reasonably high levels in nasopharyngeal swabs, throat swabs, sputum, and stool for up to two weeks[13], with the greatest amounts in saliva, sputum, stool. Saliva might be the ideal source for a home test kit, given the ease of sampling.

Tests suitable for home use are already in development. For example, an isothermal and colorimetric test has been described[14,15], based on reverse transcription-loop mediated amplification (RT-LAMP) technology. This test has several desirable properties: unlike PCR, it does not require temperature cycling; the readout is binary and can be achieved by simple observation; and it can start from crude samples[16]. Many other technologies also have the potential to detect viral RNA rapidly and isothermally[17,18]; these include recombinase polymerase amplification (RPA), transcription mediated amplification, nicking enzyme amplification reaction (NEAR), rolling circle replication, and *in vitro* viral replication assays. Finally, lateral-flow strips for the detection of viral antigen have been announced, although their performance has not yet been assessed.

The basic divide-and-conquer algorithm by which an epidemic is optimally suppressed by cordon sanitaire, quarantine and isolation was already known to the early contagionists. However, targeting of the approach is made suboptimal by uncertainty about who is and who is not infected. We have here outlined a framework for population-scale testing that will effectively decrease this uncertainty, and through rigorous analysis we have derived formulas that relate the key design parameters for the strategy. Unlike other, more complex approaches, whose optimization takes time and requires additional evidence that cannot be obtained without incurring cost in measures of population health, the societal costs of the approach we propose can be expressed in almost purely monetary terms – as a total cost of a test – which to the best of our understanding is often negative (i.e. each test leads to an increase in total economic activity). Therefore, the benefits of the proposed strategy are likely to greatly outweigh the cost, even in the event that it would fail to fully suppress the epidemic.

**Acknowledgments**

We thank many colleagues for comments on the early version of the work. We are especially grateful to Prof. Paul Romer for helpful discussions and insight, and Drs. Minna Taipale, Mikko Taipale and Paul Pharoah for reviewing early drafts of the manuscript. Earlier drafts of this paper were initially released as public preprints[19,20]. We are grateful to Nikos Demiris for a number of bibliographical pointers, and to Michail Loulakis for insightful conversations on our results. We also note that during the writing of this work, we became aware of three independent analyses, by Paul Romer (www.paulromer.net), Julian Peto[21], and by a team consisting of David Berger, Kyle Hirkenhoff and Simon Mongey[22], who report similar conclusions.

**Competing interests**

The authors declare no competing interests.

# METHODS

**Assumptions, parameters and choice of models**

The present work relies on two principal assumptions: 1) testing an individual for the presence of a pathogen will predict future infections originating from the same individual, and 2) future infections can be prevented by isolating the individual. If these two assumptions hold, an epidemic will collapse provided that the testing regime can decrease the rate of generation of new infections below the rate of recovery or death of the infected individuals. If the epidemic does not collapse, the testing regime will still decrease the rate of generation of new infections, leading to a less dramatic but still beneficial outcome. It is important to note that the above conclusions rely only on the two assumptions stated above, and not on the specific model used to assess the quantitative relationships between the testing rate and design, and summary statistic abstractions such as $R_0$.

The epidemic was first modelled with a standard (continuous, deterministic) susceptible, infected, removed (SIR) model. An important consideration in the choice of a model to represent underlying physical reality is that it represents the level of abstraction that is relevant to the scientific question addressed. For models underlying decision making, it is also very important to consider the exponential nature of the epidemic, which greatly increases the risk associated with wrong assumptions about features that are used in targeting any intervention. These include, for example, population substructure, which can lead to approaches such as TTQ to miss partially isolated subpopulations such as migrant workers. As the proposed approach is population-scale, it does not – unlike carefully targeted approaches – rely on assumptions about population substructure. Therefore, the SIR model is appropriate as an accurate first description, as it is also sufficiently abstract to capture general features that operate at a population-level, and does not fit to particular conditions in specific countries. Furthermore, the abstract nature of the model is further justified by the uncertainty in the input parameters, inclusion of each of which would require a new set of separate assumptions. Just as an example, we typically lack accurate parameters that describe the rate of test positivity as a function of future infectivity of an individual. Using any individual parameter that includes uncertainty introduces error, and using any pair of separately measured parameters that depend on each other introduces further error. Therefore, this model is solely aimed at evaluating the general feasibility of the approach, and for setting initial parameters that should be dynamically adjusted during the intervention itself.

A more general, randomized (continuous-time) Reed-Frost epidemic model on a random network was then employed, which also displays SIR-type behaviour with large population sizes. It facilitates the derivation of detailed descriptors for the epidemic as functions of the underlying parameters, leading to very similar qualitative conclusions, see below and **Supplementary Methods**.

It is important to note that parameter ranges where testing cannot effectively be used to suppress an epidemic can be found, particularly when combining worst-case estimates across studies. The weakest performance of testing is in cases with extremely high $R_0$, or very long incubation period with low viral load, followed by rapid increase and high infectiousness. The failure of some countries to suppress COVID-19, for example, by the means available to them also suggests that some initial parameter estimates are unlikely to be correct (e.g. initially it was estimated that infectious period starts after symptoms). This also indicates that a different approach is required than what is usually taken, and that asymptomatic cases need to be identified to control spread.

One important parameter that affects performance of testing is the false negative rate of tests, which varies mainly due to viral load and sampling errors. We would like to note that methods based on amplification of nucleic acids generally have higher sensitivity to detect (a correlate of) an infectious viral particle than the viral infection itself. This is due to three reasons: First, viral infection of cells is a complex process, that tends to be inefficient, with many viruses needed to establish one productive infection. Second, volume of a sample taken from a patient (microliters) is generally much higher than the total initial volume in the droplets and particles that ultimately reach the airways of the infected patients (nanoliters or picoliters[23]). Third, nucleic acid detection methods can also detect RNA released from dead cells, RNA present in dead viruses, and in the subset of defective interfering viral particles that carry the relevant segment of RNA. Despite these advantages, even PCR tests return false negatives for clinically diagnosed cases. However, the false negative rate estimated in the literature (e.g. 71% according to Ref. [24]) cannot be easily used in analysis of the effectiveness of population-scale testing. This is because the rate of false negatives is estimated from all cases, and not from infectious cases. Using two independently measured values for infectiousness and false negative rate leads to measurement in the form of a conservative lower bound for the relevant variable (time-dependent probability of obtaining a positive test result as a function of the area under the curve of future infectiousness).

**Deterministic SIR model**

In addition to the very general assumption that there are a relatively large number of cases, which allows modeling of a partially discrete system using a continuous model, the SIR model is based on the following standard assumptions: (1) the population is fixed, (2) it mixes homogenously, (3) the only way a person can leave the susceptible group is to become infected, (4) the only way a person can leave the infected group is to recover from the disease, (5) recovered persons become immune, (6) age, sex, social status, genetics etc. do not affect the probability of being infected, (7) there is no inherited immunity, and (8) the other mitigation strategies and testing are independent of each other (for **Fig. 1D**). The assumption (2) leads the SIR model to overestimate viral spread, as in reality humans react to epidemics by decreased mixing, and the population has substructure (e.g. families, workplaces), is geographically separated, and contacts are more likely between subsets of the population; these limitations are not expected to materially affect our analysis as our approach is population-scale like the SIR model itself, and the conclusions are not based on the absolute rate of the spread, only on its exponential nature. Furthermore, the SIR model is extremely conservative in the sense that it overpredicts the rate of growth of the infected population when the number of infected individuals is low; it does not even allow for extinction of the virus – a very desirable outcome, which obviously has a non-zero probability of occurrence in any real-world scenario.

In addition, we modeled the effect of testing in two ways. The first, maximally effective testing strategy assumed that every individual was tested before they infected another person, leading to the upper bound on testing performance in **Fig. 2A-B**. Under this model, the requirement for collapsing the epidemic is that the weighted average of the basic reproduction number $R_0$ and the reproduction number in isolation or quarantine $R_q$ must be less than 1.0,

$$pcR_q + (1-pc)R_0 < 1,$$

i.e.,

$$pc > \frac{R_0 - 1}{R_0 - R_q}.$$

Here, $p$ is the true positive rate of the test and $c$ is the compliance rate (fraction of all tested individuals who actually self-isolate).

Using $R_0 = 2.4$ and $R_q = 0.3$ for COVID-19 as an example, the product of the true positive rate and compliance must be greater than two thirds:

$$pc > {}^2/_3.$$

The second, lower bound testing strategy (**Fig. 2C-D**) was modelled by adding an additional 'detected' state to the model, and adding transitions from infected to detected (with rate $\tau I$) and from detected to recovered (with rate $\gamma D$). This corresponds to continuous random testing of the population at a fixed rate $\tau$ per person per day. Here, the requirement for successful collapse of the epidemic is given by the basic reproduction number (assuming perfect isolation; **Fig. 2D**), as follows. First, the rate equations for the SIR model with testing are:

$$\frac{dS}{dt} = \frac{-\beta SI}{N} \quad \frac{dI}{dt} = \frac{\beta SI}{N} - (\gamma + \tau)I$$
$$\frac{dD}{dt} = \tau I - \gamma D$$
$$\frac{dR}{dt} = \gamma(I + D).$$

Rewriting the second equation above as,

$$\frac{dI}{dt} = \left(\frac{\beta}{\gamma + \tau}\frac{S}{N} - 1\right)(\gamma + \tau)I,$$

makes it clear that $dI/dt$ will be negative (i.e. the epidemic will collapse) only if:

$$\frac{\beta}{\gamma + \tau} < \frac{N}{S}.$$

Note that the ratio $\beta/\gamma$ is the basic reproduction number $R_0$, so that the previous inequality can be rewritten as follows:

$$\tau > \gamma\left(\frac{R_0 S}{N} - 1\right).$$

In other words, the testing rate must exceed a threshold given by the recovery rate $\gamma$ (inverse of the infectious period), the time-varying reproduction number $R_0$ and the fraction of susceptible individuals $S/N$. The required testing rate drops if non-pharmaceutical interventions reduce $R_0$. Similarly, as the epidemic progresses, the required testing rate drops as fewer and fewer individuals remain susceptible and herd immunity kicks in.

**Stochastic model and analysis of the effect of testing rate**

The SIR model fails to account for several key properties of real epidemics, such as social and geographical population structure, the discrete and stochastic nature of infection and disease progression, and the fact that testing cannot be instantaneous. To account for such more complex real-world phenomena, we employed a stochastic epidemic model on a random network, wherein a contagion spreads on an arbitrary social graph, with an average infectious period of 7 days. The rigorous analysis is described in the **Supplementary methods**, and the Python code for the simulations is available on Github (https://github.com/jutaipal/contagion).

To analyze the effect of different rates of testing on the epidemic, we analyzed a stochastic Reed-Frost SIR epidemic model on a random netwrok (with respect to both the Erdős-Rényi model and the configuration model), combined with random testing. More details of the analytical framework are provided on **Supplementary Methods**. The analytical results were confirmed by numerical simulations.

**Efficiency of testing**

It is very important to know the false positive and false negative rate of a test in order to use the test optimally, and to prevent worst-case behaviour or deterministic "sign inversion" (inaccurate test that has high precision, systematically calling negative individuals positive and positive individuals negative). However, provided that the false negative and false positive rates of a test are known, they should be treated as continuous variables, and therefore a set threshold for a "good" test cannot be set independently of its intended use, as the optimal outcome can be reached using tests with widely different characteristics by varying the testing rate. To further understand the efficiency of testing, it is helpful to consider three separate processes: the process of the infection itself, the ordering of the tests relative to the infectious process, and the structuring of the testing process.

The that an individual carries a virus can be divided into four consecutive intervals: undetectable (u), lead time for testing (l), infectious period (i) and antigen positive but non-infectious period (a). In a model of testing, the classical incubation period has to be split into two (u+l), because the relevant interval for antigen/genetic testing to suppress an epidemic by isolation of the tested individuals is not the infectious period i, but l+i; l exists because of the very large difference in volume transfer between a test and an infection in the case, .e.g, of a respiratory virus. The period relevant for contact-tracing is even longer, l+i+a, as detection of the recent infection leads to identification of a starting point for the search of currently infected contacts. It is important also to note that the intervals are abstractions of infectious and non-infectious virus genetic material / particle concentration over time within an individual; this is never 0 in an individual who carries the virus, and varies in both time and magnitude, affected by the precise dynamics of the infection process itself.

To understand the difference between testing (1) the vector, (2) prior to transmission, (3) everyone at the same time, (4) everyone in a time separated manner, or (5) the population by random sampling, it is helpful to consider the extreme case of certainly and completely collapsing an epidemic by testing and isolation, using a perfect test that detects all infected individuals and complete isolation. For optimally achieving this, it is necessary to identify everyone who is infected before they have infected anyone else (denoted efficiency, $e = 1$) and quarantining every infected individual ($c = 1$). This requires obtaining a minimum of $n$ bits of information for a population of size $n$. Note, however, that a single test performed as a group test can return more than 1 bit, and that this can be useful in detection of super-spreaders with very high viral titer (Ref. [25]). The order of performance of different strategies can be derived from the following simple model:

Let there be a constant number of tests (**T**) per individual placed on sequences consisting of four elements: $I_0$ (incoming infection of individual 0), **T** (test), $I_n$ (infection of individual n, with I without index indicating any infection) and R (recovery of the individual).

All sequences in case (1) have the following form:
$$\textbf{T}\text{-}I_0\text{-}(I \text{ or } R).$$
This allows the test to prevent even the first infection, which is clearly the best possible way to place a single test per individual.

All sequences in case (2) have the following form:
$$I_0\text{-}\textbf{T}\text{-}(I \text{ or } R).$$

This cannot prevent the first infection, but is clearly the second-best possible way to place a single test per individual.

Case (3) allows the second-best order,

$$I_0\text{-}\textbf{T}\text{-}(I \text{ or } R),$$

but also recursively:

$$I_0\text{-}I_1\text{- } \ldots \text{ -}I_n \text{ -}\textbf{T}\text{-}(I \text{ or } R) \quad I_1\text{-}I_{1,1}\text{-}\ldots \text{ -}I_{1,n}\text{-}\textbf{T}\text{-}(I \text{ or } R)$$
$$I_{1,1}\text{-}\ldots\text{-}I_{1,1,n}\text{-}\textbf{T}\text{-}(I \text{ or } R)$$
$$\ldots$$
$$I_n\text{-}\textbf{T}\text{-}(I \text{ or } R).$$

Note that the concurrency ensures that tests are placed on each interval optimally (one on each branch of the tree).

Whereas case (4) can allow the above sequences, and in the absence of specific ordering of infections and tests,
$$I_0\text{-}\textbf{T}\text{-}I_1\text{-}(I \text{ or } R)$$
$$I_1\text{-}I_{1,1}\text{-}\ldots \text{ -}I_{1,n}\text{-}\textbf{T}\text{-}(I \text{ or } R),$$
which places two tests along the same chain, and also recursively,
$$I_0\text{-}I_1\text{- } \ldots \text{ -}I_n \text{ R}$$
$$I_1\text{-}I_{1,1}\text{-}\ldots \text{ -}I_{1,n} \text{ R},$$

a branch with no tests at all. Ordering of tests and infections, by, for example, sweeping a region and preventing movement of individuals in the opposite direction improves this strategy, but its performance will remain below case (3) because the branches will grow until a test is placed on them.

Case (5) performs the worst, because it also allows,

$$I_0\text{-}\mathbf{T}\text{-}\mathbf{T}\text{-}I_1,$$

which tests the same individual twice during one infectious period. This is in itself good, but decreases overall efficacy of the regime as it leads to more branches with no tests than case (4).

Case (3) is the weakest testing regime that always achieves complete collapse; this is achieved by testing everyone at the same time with a perfectly accurate test that returns one bit (positive or negative). In this case, $e = 1$ and $c = 1$. However, when tests are separated in time (Case 4), the order of testing becomes important. Most strategies for testing n individuals during time $t_{\text{test\_interval}}$ before $t_0$ have $e < 1$, and are not sufficient to completely collapse the epidemic using one testing round, as $e$ depends on the relationship between the order of testing and the order of infections. For example, using a random order of testing allows some individuals that have already been tested negative to become infected during the $t_{\text{test\_interval}}$ (the mutual information between test results and person being infected at $t_0$ is less than one bit). However, some other regimens using a perfectly sensitive test can collapse the epidemic (but not always prevent all future infections): for example, a geographical sweep where infections (individuals) are prevented from crossing a moving test front can be used to identify every infected individual in the population by performing a single round of n tests.

In case (5), random sampling of n individuals, $e$ is always less than 1. The testing becomes less efficient than testing each of the n individuals at the same time, because some individuals are tested twice, and some not at all; some information is thus not obtained, and some tests do not return information that is completely independent of information returned by other tests (sum of mutual information between all pairs of tests is not 0 bits). In other words, if individuals are selected randomly, during a given time interval, the tests will miss some individuals, and some individuals are tested more than once (this increases true positive rate for those individuals, but this does not make up for failing to catch some individuals entirely).

Considering the extreme case of immediate collapse, it may appear that testing in a time separated manner or by using random sampling will not work because non-concurrent testing can permit infections to cross the testing boundary, and random sampling clearly leaves some cases undetected. However, this very intuitive idea is incorrect, as collapsing an epidemic only requires that the rate of generation of new cases per current case is less than one. The limit for random testing can be obtained using the SIR model extended with testing (SIR+T), which abstracts away individuals and thus can (only) be used to investigate the effect of random, time-separated testing. Analytically from this model, as shown above, the $R < 1$ condition is true when tests are performed at a rate that is higher than $R_0 - 1$ tests per mean infectious period. The same limit results from the consideration of a model that is similar to the the simple model above: reducing $R_0$ to less than 1.0 using the method representing the lower bound – a completely random testing regime – requires that an infected individual has less than an equal probability of (a) infecting another individual over (b) being tested and isolated or recovering from the infection (analogously, in SIR+T, the combined testing and recovery rate needs to be higher than the rate of new infections). Events (a) can recur, but either event (b) terminates the chain. Therefore, at R = 1 there will be on average one (a) event, which requires that the order of the infectious and protective events are randomly ordered with respect to each other, with equal density. This yields $R_0 - 1$ tests and one recovery per $R_0$ infections per infectious period, and an upper limit of $R_0$ tests per infectious period at infinity (because as $R_0 \to \infty$ the expected value for the number of test required per $R_0$ becomes the geometric series $\sum_{n=1}^{\infty} 2^{-n} = 1$).

Outside of the theoretical consideration $e = 1$, multiple population-scale tests are always required to collapse the epidemic in the absence of other interventions that achieve the same aim. Performing multiple tests over time imposes an additional constraint on optimality – the allocation of tests to each transmission interval. As described above, best performance of continuous testing and isolation is thus achieved when testing is performed immediately after infection for each individual, or as requirement for exiting quarantine. Testing at border crossings, conditional opening of lockdown, or some regimes that apply contact-tracing may come close to approximating this limit, which for COVID-19 is $pc = (R_0 - 1)/R_0 = 0.57$ per mean infectious period; **Fig. 2**). However, in most scenarios, such testing efficacy is difficult to maintain over time (because contact is lost, and the unknown infectious intervals rapidly become randomly distributed over time). This level can thus be considered an upper limit of performance of any scenario applied at population scale.

Using a test whose true positive rate is 1 and testing everyone at the same time performs as well as the optimal strategy. As test sensitivity decreases, the performance of the concurrent regime becomes lower than optimal. However, concurrent testing still performs well above the lower limit obtained from the random testing model. The required pc rate to bring $R_0 < 1$ using concurrent tests has a simple relationship with the exponential growth of infectious cases. Over interval t-$t_0$, pc > 1-(infectious cases at $t_0$)/(infectious cases at t). However, it is not as simple to relate this to original $R_0$, because the relationship between $R_0$ and growth rate is a function of the distribution of the generation intervals (Ref. [26]). Estimating at $R_0 = 2.4$ using even probability distribution of infections over time, the infected population becomes approx. eight times larger in a single infectious period (for SIR, this would be different due to the different assumption about the distribution of infectious periods; see Ref. [26]). This means that a testing regime that is regularly spaced and once per infectious period should have pc value of > 7/8 = 0.875 to bring $R_0 < 1$. This is confirmed using empirical simulations to assess the rate of exponential growth in the complete absence of immunity and all other types of interventions; the limit $R = 1$ at $R_0 = 2.35$ with testing every infectious period is reached when $pc \approx 0.85$ (compared to 0.58 for testing each individual directly after infection). The required testing interval at $R_0 = 2.35$ and $pc = 0.8$ in the absence of other interventions and immunity is ~ 0.8, 0.6 and 0.4 times the infectious period for concurrent testing, testing each individual randomly once during each testing period, and continuous random testing, respectively.

These considerations can be summarized as follows: the order of testing efficacies is: all vectors (at edges of network) > everyone before they have had a chance to infect anyone > everyone at the same time > everyone once during a period > testing by random sampling – with population-scale testing remaining feasible and cost-effective by one or more orders of magnitude across all these regimens.

For the structuring of the testing process, one should carefully consider the entire testing regime, which is invariably population-scale for all infectious diseases, even if this intuitively may appear not to be the case. In practice, many different approaches are used, but most behave formally as consecutive testing, with the increased number of layers commonly (optimally in any real situation) increasing both the true positive and the false negative rate. This is because the first test (e.g. appearance of symptoms, or being a contact or not) can increase the true positive rate, but can also determine a lower bound for the false negative rate of the whole process (e.g. in a purely contact-tracing based approach, infection of a missed contact cannot be detected biochemically). Thus, it is important to understand that while predictive tools such as appearance of symptoms or contact-tracing can increase the true positive rate, this invariably comes at some cost of increase in the false negative rate (e.g. asymptomatic individuals, lost contacts). Many predictive approaches are also highly time-dependent; for example, timing of tests in contact-tracing optimally approaches the best case (testing before infectivity), but in the worst case performs far worse than random testing (testing after an individual has already infected everyone they can). Thus, due to the increase in false negative rate, a contact-tracing approach must both be combined with another mechanism to catch the escaped contacts, and performed extremely fast to improve the timing of the tests. A fully optimal design is likely to incorporate multiple different strategies, including social distancing, contact tracing, and symptom-based and population-scale testing. In designing a practical strategy, however, it is important to understand that two of the most important variables to optimize are the amount of potential primary evidence (infected individuals), and time. Given that the mathematics involved are non-trivial, arriving at an optimal solution will take considerable time. Therefore, action before arriving at an optimum is critical. In other words, one should consider the relative costs of each approach, and the difficulty of rapidly implementing complex and time-dependent processes. Furthermore, regardless of its level of sophistication, any layered or consecutive approach will always – by definition – have both a complexity and a time-cost that population-scale biochemical testing approach will eliminate.

# FIGURES

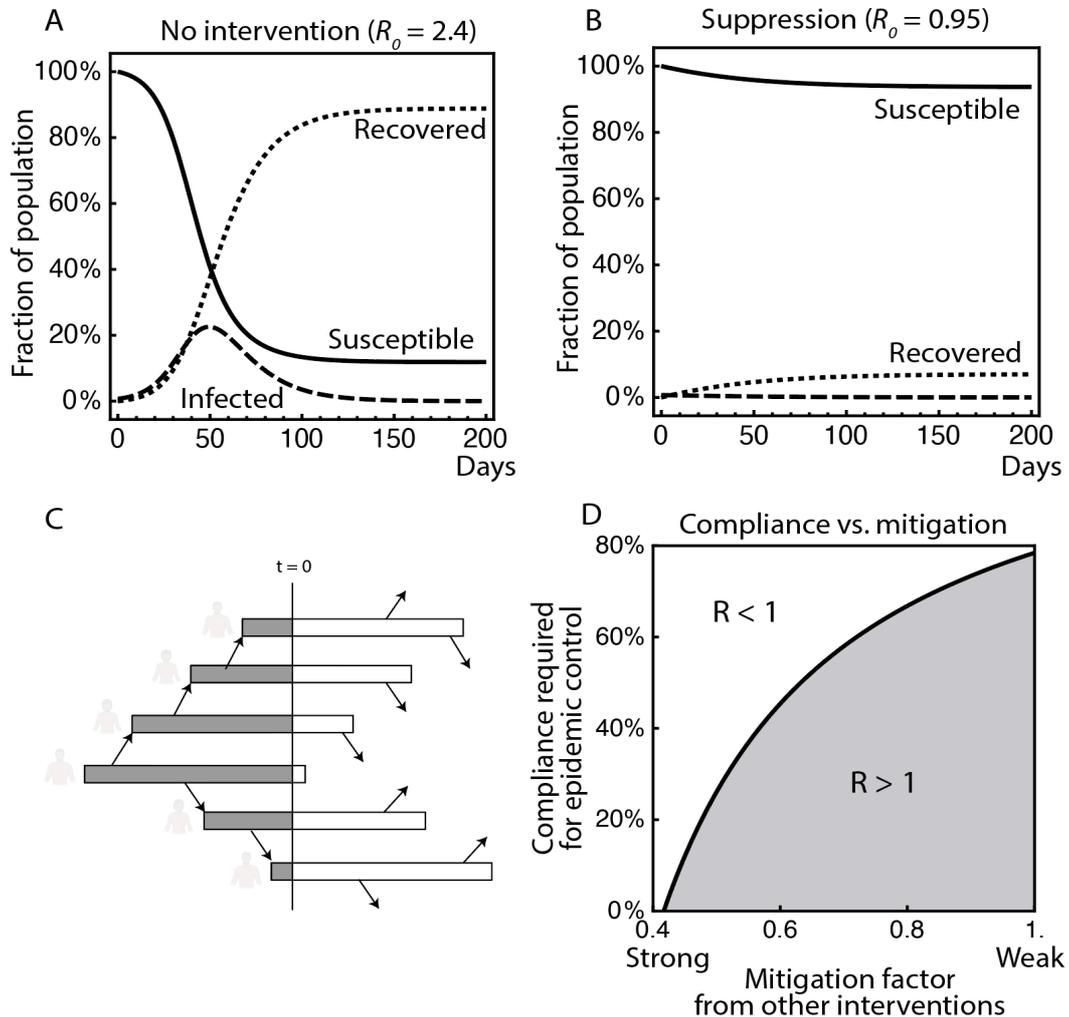

**Fig. 1** | Requirements for epidemic control using population-scale testing and self-isolation.
(A) An SIR model (Ref. [27]) with $R_0 = 2.4$ leads to infection of the majority of the population, with a massive peak of active infection that overwhelms the health care system.
(B) With strong interventions that reduce $R_0$ to 0.95, many deaths are avoided.
(C) Testing everyone simultaneously cuts all chains of transmission.
(D) The required level of compliance (as a fraction of all individuals) for effective control of the epidemic, as a function of the strength of other interventions and assuming a test with 85% efficacy (fraction of future infections detected). With moderate social distancing, epidemic control can be achieved even with low levels of test compliance.

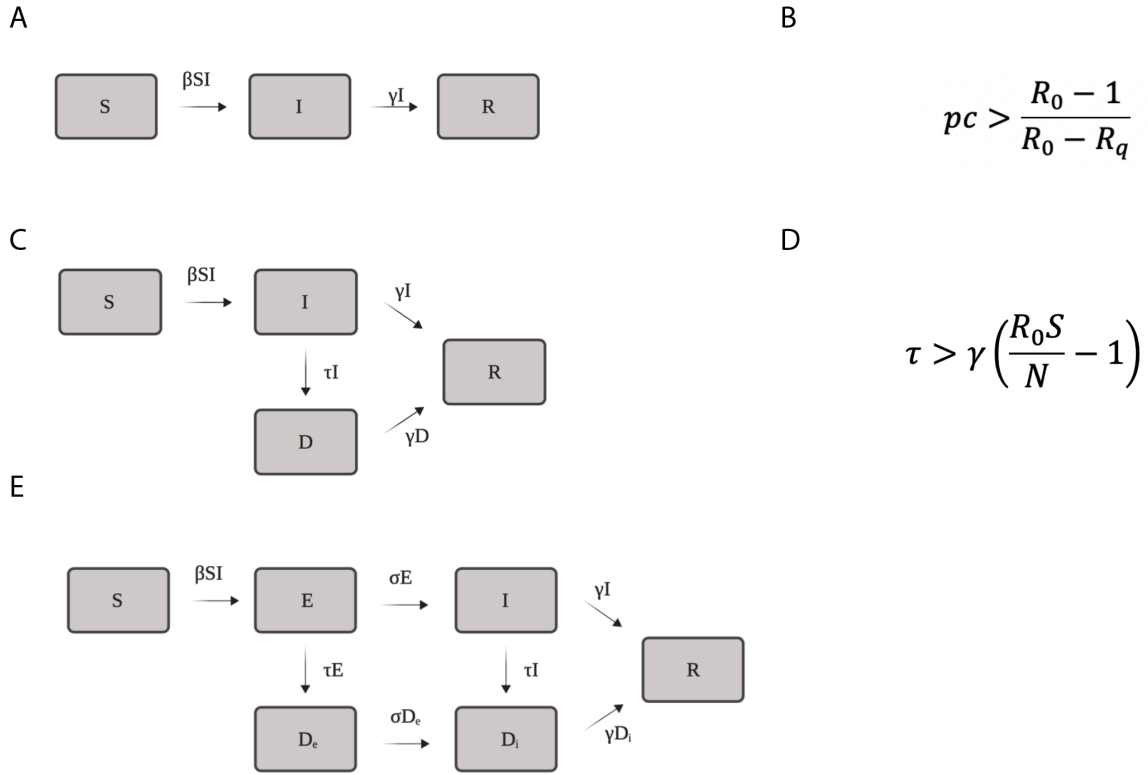

**Fig. 2 | Compartment models and requirements for suppression by testing.**

(A) Parameters for a standard SIR model.

(B) Inequality that must be true to suppress transmission. For the epidemic to collapse, the weighted average of the natural reproduction number $R_0$ and the reproduction number in self-quarantine $R_q$ must be less than 1.0. Here, $p$ represents the test true positive rate (fraction of all infectious individuals detected), and $c$ the rate of compliance.

(C) Parameters for a SIR model with testing and a detected state.

(D) Requirements for testing to collapse an epidemic in the SIR model with testing, expressed in terms of the testing rate $t$ required in a population where all individuals are susceptible, with inverse infectious interval $\lambda$.

(E) Parameters for the discrete, stochastic SEIR model.

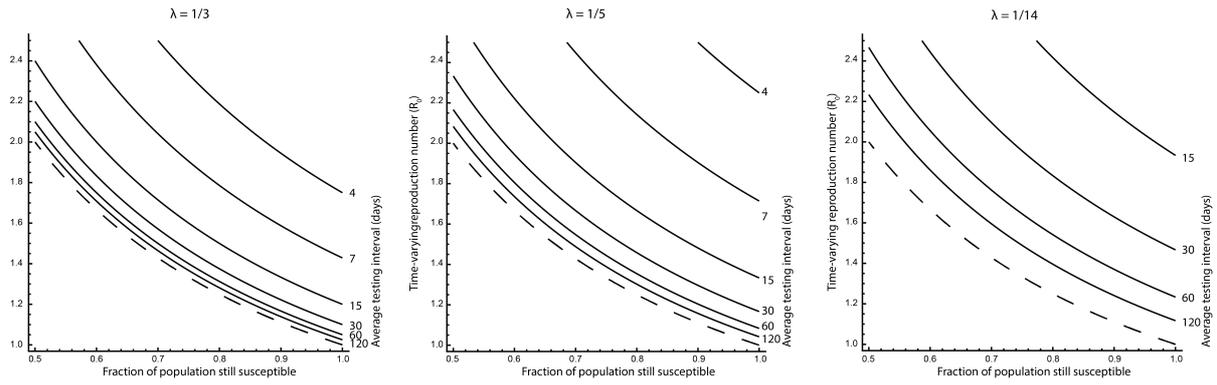

**Fig. 3** | Testing frequency required to control transmission as a function of the fraction still susceptible ($S/N$, horizontal axes) and the reproduction number ($R_0$, vertical axes) for three recovery rates corresponding to 3, 5 and 14 day infectious periods. Solid curves are testing frequencies, indicated as the average number of days between tests, per person. Dashed curves depict the herd immunity threshold below which transmission collapses without testing.

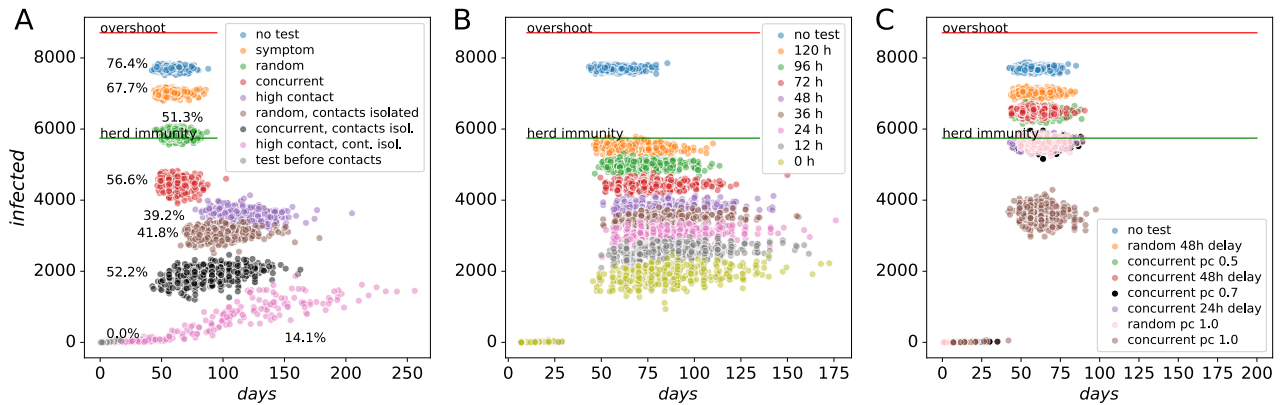

**Fig. 4 | Controlling an epidemic using tests requires a very rapid test employed at population scale.**

(A) Testing individuals before contacts (gray), concurrent testing of all individuals (red, black), or testing individuals with a high number of contacts (violet, pink) outperforms random population-scale test-and-isolation regimes (green, brown). Isolation of direct contacts of infected individuals (brown, black, pink) improves the respective regimens. All the high-performing regimes require very rapid action or very high numbers of tests per high-contact individual. To compare between the regimes, the epidemiological parameters were set so that the effect of testing is on the "linear range" with respect to the final size of the epidemic ($y$-axis). The example shows results when each individual is on average tested once per 7-day infectious period; population = 10,000, one infection at $t = 0$, $R_0 = 2.35$, 20 contacts on average with a "scale-free" distribution (on a Barabasi-Albert-type graph). Percentages indicate fraction of simulations with >100 infected. In the "high contact" testing regime, the probability of testing increases approximately linearly with the number of contacts up to a maximum of twice per day.

(B) Delays in reporting test results severely impact the ability to control an epidemic by testing. The simulated regime is concurrent testing with contacts isolated. See also Ref. [28].

(C) Relationship between delay, concurrency and pc = (true positive rate x isolation compliance). Note that even a 48-hour delay in reporting is equivalent to a decrease of the true positive rate from 1.0 to 0.5, and that random testing at 1.0 pc and no delay performs as well as concurrent tests with a 24-hour delay, or concurrent tests at 0.7 pc without delay. The specific numbers depend on the input parameters, but the indicated order of performance is retained.

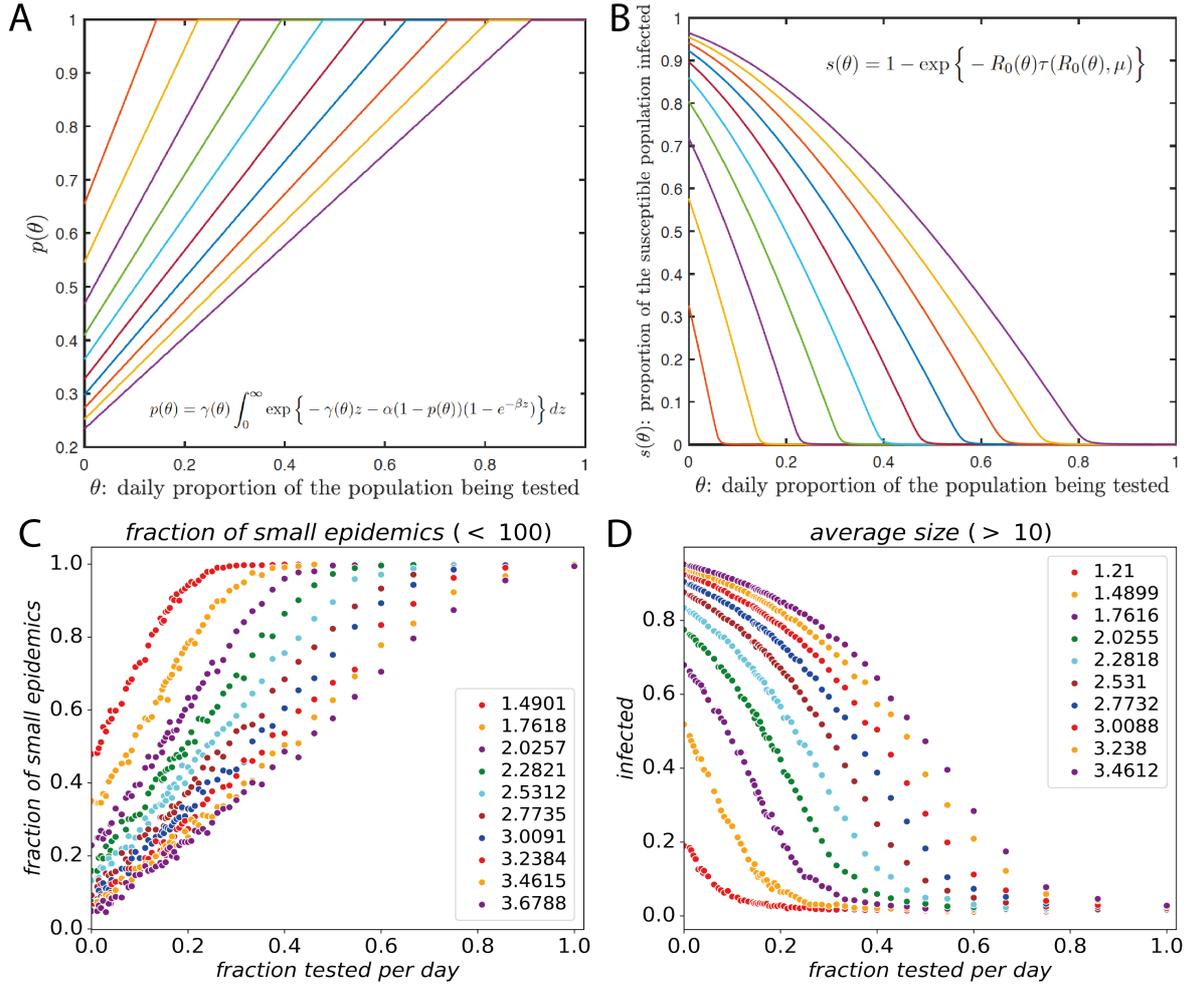

**Fig. 5** | Effect of testing rate on the epidemic.

(A) Increasing the rate of random testing of the population linearly increases the probability of early extinction (denoted $p(\theta)$) of a virus introduced to a population by a single infected individual. The equation that describes the relationship between $p(\theta)$ and the fraction $\theta$ of population tested daily is also shown. The figure shows results for a series of $R_0$ values at pc = 0.525 with a 7 day infectious period. For details, see **Supplementary Methods**.

(B) Increasing the rate of random testing of the population decreases the final size of the epidemic $s(\theta)$. Note that any increase in the rate of testing is beneficial, even when the testing rate is lower than the critical rate that leads to the complete collapse of the epidemic.

(C) Numerical simulation of an epidemic (see **Methods**) using the same $R_0$ values (inset) as in panel (A) confirms the linear relationship between testing rate and the probability of extinction. For the purpose of the simulation, a small epidemic is considered one with less than 100 infected individuals prior to extinction of the virus introduced to the population by a single infected individual.

(D) Numerical simulation of the final size of epidemics using the same $R_0$ values (inset) as in panel (B). Note that the simulation confirms the overall prediction of the analytical solution (**Supplementary Methods**) at higher $R_0$ values, when the number of infected individuals becomes either very small or very large. The numerical simulation predicts smaller size of epidemics that affect more than 10 individuals than the analytical result (B) at low $R_0$. This effect is caused by the fact that, at low $R_0$, epidemics can also commonly collapse after expanding beyond 10 individuals, but before reaching the larger final sizes, whose average size is predicted by the analytical solution in (B).

# SUPPLEMENTARY METHODS

## 1 Erdős-Rényi network

### 1.1 Model and parameters

We first describe the randomized Reed-Frost epidemic in continuous time, on a Bernoulli random graph; see [10, 2, 8] and the references therein.

Consider a population initially consisting of $a$ infected individuals and $N$ susceptible ones. In this discussion we think of $N$ as being large, so we also describe how each of the parameters involved should be appropriately scaled. For the initial infection we assume,

$$a = a_N = \# \text{ of initially infected individuals} = \mu N,$$

for some fixed, typically small $\mu > 0$. Every individual is either connected or not connected with every other individual, with probability,

$$p = p_N = \text{connection probability} = \frac{\alpha}{N},$$

independently among different pairs, for some $\alpha > 0$. This corresponds to a sparse Erdős-Rényi random graph [5, 4]. We call connected individuals *acquaintances*. For large $N$, each individual has an approximately Poisson distributed number of acquaintances, with mean,

$$\alpha(\mu + 1) = \text{mean } \# \text{ of an individual's acquaintances}.$$

An infected individual, say individual $i$, stays infected for an $\text{Exp}(\gamma)$ amount of time, independently between individuals and of the graph, after which she becomes recovered. During her infected phase she has infectious contacts with some of her $d_i$, say, acquaintances, at the times of a Poisson process with rate $\beta d_i$. These Poisson processes are all independent, independent of the infection durations and of the random graph. The rates $\beta > 0, \gamma > 0$ are fixed:

$$\frac{1}{\gamma} = \text{mean infection duration}, \qquad \beta = \text{rate of individual infectious contacts}.$$

At each of these Poisson times, she has an infectious contact with a uniformly and independently chosen individual among her $d_i$ acquaintances. If the contacted individual is susceptible, they become infected; otherwise, nothing changes.

From the results of [2] and their interpretation in [8], we know that the resulting SIR model describes an epidemic with basic reproduction number,

$$\boxed{R_0 = \frac{\alpha \beta}{\beta + \gamma}} \tag{1}$$

Now let us assume that, in parallel with the infection's dynamics, an independent test-and-quarantine process is in place. At the event times of an independent Poisson process with rate $r > 0$, an individual is selected uniformly at random and given a test with *sensitivity* $1 - \delta$, i.e., with false-negative error rate,

$$\delta = \text{false-negative probability} = \Pr(\text{negative test result}|\text{infected}),$$

for some $\delta \in [0, 1]$.

We ignore false positives, or, equivalently, we assume that the test has perfect specificity, i.e., it never produces a false positive result. Note that this can only make our subsequent predictions on how much of the population needs to be tested per day in order for the epidemic to be contained, more pessimistic, that is, we may only overestimate how much testing is necessary.

We assume that the testing rate is proportional to the population size,

$$r = r_N = \text{testing rate} = \rho N,$$

for a fixed $\rho > 0$. If an individual's test is positive, they are asked to stay in quarantine, in effect becoming recovered. They comply with this request with,

$$q = \text{probability of quarantine compliance},$$

for some fixed $q \in [0,1]$, independently of all other variables. If the test is negative, nothing happens.

## 1.2 Reproduction number and epidemic size

**Effect of testing.** We claim that the test-and-quarantine protocol only affects the parameter $\gamma$ in the original model. First of all, quarantine non-compliance and test errors can be viewed as simply "thinning" the testing process, and the model is equivalent to a perfect test and full compliance with a testing rate,

$$\rho' N := [\rho(1-\delta)q]N.$$

Then, the actual duration of an infected individual's infection period is distributed as $Z' = \min\{Z, W\}$, where $Z \sim \text{Exp}(\gamma)$ and $W \sim \text{Exp}(\rho'/(\mu+1))$ are independent. This is also an exponential random variable with parameter,

$$\gamma' = \gamma + \frac{\rho'}{\mu+1} = \gamma + \frac{\rho(1-\delta)q}{\mu+1}. \tag{2}$$

The resulting epidemic is identical to the original one, except that now the mean infection duration is $1/\gamma'$ rather than $1/\gamma$.

**Reproduction number.** The test-and-quarantine procedure decreases the value of $R_0$ given in (1) to $R'_0$, where,

$$\begin{aligned} R'_0 &= \frac{\alpha\beta}{\beta+\gamma'} \\ &= \frac{1}{1+\frac{\rho q(1-\delta)}{(\mu+1)(\beta+\gamma)}} \times R_0. \end{aligned}$$

To express this in a way that is easier to interpret, suppose that the time units are days. Then randomly testing a proportion $\theta$ of the population every day, so that $\theta(\mu+1)N = \rho N$, i.e.,

$$\boxed{\theta = \frac{\rho}{\mu+1}}$$

corresponding to $\rho = \theta(\mu+1)$, yields a basic reproduction number,

$$\boxed{R_0(\theta) = \frac{\alpha\beta}{\beta+\gamma+\theta(1-\delta)q}} \tag{3}$$

An important observation is that $R_0(\theta)$ falls below 1 as soon as $\theta$ becomes larger than a critical value $\theta^*$, where,

$$\boxed{\theta^* = \frac{\alpha\beta - \beta - \gamma}{q(1-\delta)}}$$

Simple numerical illustrations of the above quantities are given in Section 1.4.

**Total size.** Testing also reduces the total size $T_N$ of the epidemic, where $T_N$ denotes the number of the initially susceptible individuals that became infected at some point. Following the exposition in [8], we let $\omega = \alpha E[1 - e^{-\beta Z}]$, where $Z \sim \text{Exp}(\gamma)$, so that, $\omega = R_0 = \alpha\beta/(\gamma+\beta)$. And with,

$$\tau = \tau(\omega, \mu) := \min\{t > 0 \; : \; e^{-\omega t} = 1 + \mu - t\}, \tag{4}$$

we define $s = 1 - e^{-\omega\tau}$. The first part of following result can be found in [8]; the second part follows form the first part combined with the argument in the beginning of this section.

**Theorem 1.1** (i) **No testing.** *The total size $T_N$ of the epidemic, as $N \to \infty$, satisfies,*

$$\frac{T_N - sN}{\sqrt{N}} \xrightarrow{\mathcal{D}} N(0, \sigma^2),$$

*where the variance $\sigma^2$ can be explicitly identified. In particular, the proportion of the population that eventually has been infected,*

$$\frac{T_N}{N} \to s, \qquad \text{in probability, as } N \to \infty.$$

(ii) **Random testing.** *Under the above test-and-quarantine protocol, the same result holds, except now the mean and variance are different. In particular, $s$ now is given by $s' = 1 - e^{-R_0'\tau'}$, where $\tau' = \min\{t > 0 \; : \; e^{-R_0' t} = 1 + \mu - t\}$.*

To see the effect of testing on $s$, suppose again that a random proportion $\theta$ of the population is tested daily, so that $\rho = \theta(\mu + 1)$ and $R_0(\theta)$ is given by (3). Then the proportion of the population that will have been infected eventually is,

$$\boxed{s(\theta) = 1 - \exp\Big\{ - R_0(\theta)\tau(R_0(\theta), \mu)\Big\}}$$

where the function $\tau$ is defined in (4); see Section 1.4 for examples with explicit numerical illustrations.

## 1.3 Small epidemics

Next we consider an epidemic with a *fixed* number $a_N = a$ of initially infected individuals. We keep the scaling of all the parameters except $a_N$ to be the same as before, but we assume that $a_N = a$ stays constant with $N$, leading to a limiting value of $\mu = \lim_{N \to \infty} a_N/N = 0$. Define,

$$\begin{aligned} f(t) &= E\Big[\exp\Big\{ - \alpha(1-t)(1 - e^{-\beta Z})\Big\}\Big] \\ &= \gamma \int_0^\infty \exp\Big\{ - \gamma z - \alpha(1-t)(1 - e^{-\beta z})\Big\} \, dz, \end{aligned}$$

where $Z \sim \text{Exp}(\gamma)$, and write $p$ for the smallest root of $p = f(p)$ in $[0, 1]$. Let $R_0 = \alpha\beta/(\beta + \gamma)$ as before.

The following theorem describes the size of the resulting epidemic as $N \to \infty$. The first (no-testing) part can be found in [7, Theorem 1]; the second part follows from the first part combined with the argument in the beginning of Section 1.2.

**Theorem 1.2** (*i*) **No testing.**

- (a) *Suppose $R_0 > 1$. With probability $p^a$ there is only a **small epidemic**, that is, the epidemic dies out and its total size $T_N$ converges in distribution to a finite limit $T$; or, with probability $1 - p^a$ there is a large epidemic, $T_N$ grows linearly with $N$, and its limiting behaviour is the same as in part (i) of Theorem 1.1.*
- (b) *Suppose $R_0 < 1$. Then $p = 1$ and we always have a small epidemic, so that only the first piece of the limit distribution is present.*

(*ii*) **Random testing.** *Assume there is a test-and-quarantine protocol in place, with parameters $\delta$, $\rho$ and $q$. Let $R'_0 = \alpha\beta/(\beta + \gamma')$, with $\gamma'$ as in (2), and let $p'$ be defined like $p$ but with $\gamma'$ instead of $\gamma$ in the definition of $f(t)$.*

- (a) *Suppose $R'_0 > 1$. With probability $(p')^a$ there is only a **small epidemic**, that is, the epidemic dies out and $T_N$ converges in distribution to a finite limit $T'$; or, with probability $1 - (p')^a$ there is a large epidemic and the limiting behaviour of $T_N$ is the same as in part (ii) of Theorem 1.1.*
- (b) *Suppose $R'_0 < 1$. Then $p' = 1$ and we always have a small epidemic, so that only the first piece of the limit distribution is present.*

Suppose, again, that a proportion $\theta = \rho$ of the population (on the average) is randomly tested every day. Then the *probability of a small epidemic* is $p(\theta)^a$, where, $p(\theta)$ is the smallest root in $[0, 1]$ of the equation,

$$p(\theta) = \gamma(\theta) \int_0^\infty \exp\left\{-\gamma(\theta)z - \alpha(1 - p(\theta))(1 - e^{-\beta z})\right\} dz, \qquad (5)$$

where,

$$\gamma(\theta) = \gamma + \theta(1 - \delta)q.$$

This is illustrated via numerical examples in the following Section 1.4.

### 1.4 Summary and numerics

The table below summarizes all the epidemic model parameters, with and without testing. In addition to these, the probability $p(\theta)$ of a small epidemic when starting with a single infected individual, is given as the solution of equation (5) above.

Suppose that the time units are days. In the numerical examples in this section we assume that 1-in-10,000 individuals are initially infected, so that $\mu = 1/9999$, having an average of $\alpha(\mu + 1) = 20$ acquaintances each, so that $\alpha = 19.998$. The infectious period is assumed to last for 7 days on the average so that $\gamma = 1/7$, and we let the contact rate $\beta$ between individuals vary, corresponding to different scenarios with different levels of social distancing policies.

|   | **Basic Parameters** |   |
|---|---|---|
| $N$ | initial susceptibles | $N$ = number of initially susceptible individuals |
| $\mu$ | initial infected | $\mu N$ = number of initially infected individuals |
| $\alpha$ | connectivity | $\alpha(\mu+1)$ = mean number of acquaintances/individual |
| $\beta$ | contact rate | $\alpha\beta(\mu+1)$ = mean number individual contacts/day |
| $\gamma$ | infection duration | $1/\gamma$ = mean length of infection period |
| $q$ | quarantine compliance | $q$ = probability of compliance |
| $\delta$ | test sensitivity | $\delta$ = probability of false negative test |
| $\rho$ | testing rate | $\rho N$ = mean number of tests per day |
|   | **Derivative Parameters** |   |
| $R_0$ | basic reproduction number | $R_0 = \frac{\alpha\beta}{\beta+\gamma}$ |
| $\theta$ | daily tests | $\theta = \frac{\rho}{\mu+1}$ = proportion of the population tested daily |
| $R_0(\theta)$ | $R_0$ with testing | $R_0(\theta) = \frac{\alpha\beta}{\beta+\gamma+\theta q(1-\delta)}$ |
| $\theta^*$ | critical testing rate | $\theta > \theta^* = \frac{\alpha\beta-\beta-\gamma}{q(1-\delta)} \Rightarrow R_0(\theta) < 1$ |
| $s(\theta)$ | epidemic size | $s(\theta) = 1 - \exp\{-R_0(\theta)\tau(R_0(\theta),\mu)\}$, cf. (4) |

The sensitivity of the test is taken to be 70% so that $\delta = 0.3$, and quarantine compliance is assumed to be 75% so that $q = 0.75$. The values of all these parameters are the same in the results of all four figures shown next, except in the last case where we assume that $\mu = 0$, $\alpha = 20$ and only one infected individual is present initially.

Figure 1 shows the effect of testing on $R_0$ in ten different cases. Note that each curve intersects the horizontal line $R_0 = 1$ at the critical value $\theta^*$.

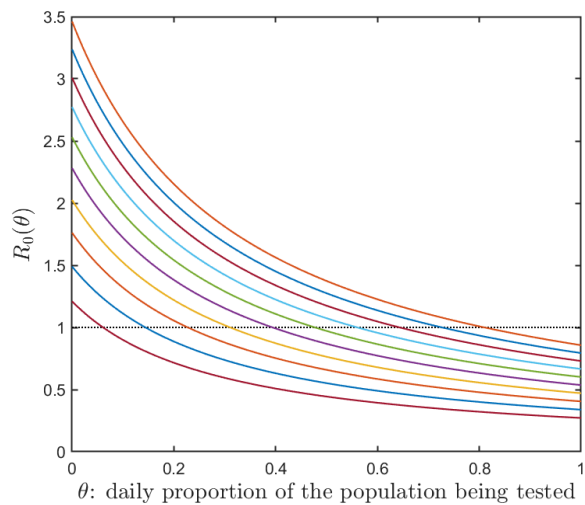

Figure 1: Effect of randomly testing a proportion $\theta$ of the population daily on the basic reproduction number $R_0$, in ten different cases, with contact rates $\beta = 0.0092, 0.0115, 0.0138, 0.0161, 0.0184, 0.0207, 0.0230, 0.0253, 0.0276, 0.0299$, and corresponding values for $R_0 = 1.2100, 1.4899, 1.7616, 2.0255, 2.2818, 2.5310, 2.7732, 3.0088, 3.2380, 3.4612$.

Figure 2 shows the proportion $\theta = \theta^*$ of the population that needs to be tested daily in order for $R_0(\theta)$ to drop below 1, for different scenarios corresponding to different values of $R_0$. Observe that $\theta^*$ can be greater than 1, as some individual may be tested several times on the same day.

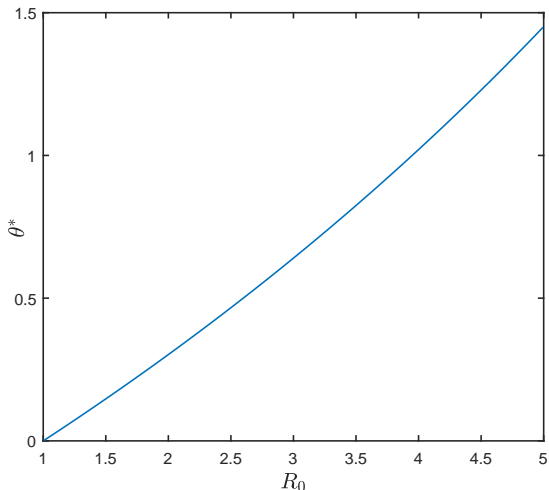

Figure 2: The critical testing rate $\theta^*$ for different scenarios with different values of $R_0$. The contact rate $\beta$ varies between 0.0076 and 0.0476, corresponding to values of $R_0$ between and 1 and 5, giving $\theta^*$ between 0 and 1.4494.

Figure 5B in the main text shows the total size $s(\theta)$ of the epidemic (i.e., the proportion of susceptible individuals that eventually have been infected), as a function of the testing rate, in ten different cases, with contact rates $\beta = 0.0092, 0.0115, 0.0138, 0.0161, 0.0184, 0.0207, 0.0230, 0.0253, 0.0276, 0.0299$, and corresponding values for $R_0 = 1.2100, 1.4899, 1.7616, 2.0255, 2.2818, 2.5310, 2.7732, 3.0088, 3.2380, 3.4612$.

And Figure 5A in the main text shows the probability $p(\theta)$ of a small epidemic, when the infection starts with a single infected individual, as a function of the daily testing rate, in ten different cases, with contact rates $\beta = 0.0115, 0.0138, 0.0161, 0.0184, 0.0207, 0.0230, 0.0253, 0.0276, 0.0299, 0.0322$, and corresponding values for $R_0 = 1.4901, 1.7618, 2.0257, 2.2821, 2.5312, 2.7735, 3.0091, 3.2384, 3.4615, 3.6788$.

## 2 Epidemics on the configuration model

Here we examine the evolution of a randomized Reed-Frost-type epidemic on the more general class of random graphs described by the *configuration model* introduced by Bollobás [4]; see also [11]. Our development follows the analysis in [6]; closely related results can also be found in [1, 2, 9, 3].

We consider an epidemic evolving exactly as before, but this time the underlying graph is a randomly drawn sample among all graphs with a given degree distribution. Specifically, given a distribution $\{p_k \ ; \ k \geq 0\}$ on the nonnegative integers, our graph will be uniformly drawn among all graphs with $N$ vertices such that the proportion of vertices with degree $k$ is approximately equal to $p_k$, for each $k \geq 0$. Further mathematical details are given in Section 2.3.

### 2.1 Basic reproduction number

Given a graph consisting of $N$ individuals, drawn from the model described above with degree distribution $\{p_k\}$, the epidemic evolves as follows. Initially there are $\mu N$ infected individuals and $(1-\mu)N$ susceptible ones, an infected individual stays infected for an Exp($\gamma$) amount of time, and while infected she has infectious contacts with some of her $d_i$ acquaintances at the times of a Poisson process with rate $\beta d_i$. At each of these event times, she has an infectious contact with one of her acquaintances selected at random. We assume that there is a parallel testing-and-quarantine process, with parameters $\delta, \rho$ and $q$, defined exactly as before.

In the following theorem we describe the basic reproduction number of this epidemic, with and without testing. Let $\lambda$ denote the mean of $\{p_k\}$, $\lambda = \sum_{k=0}^{\infty} k p_k$, and write $v^2$ for the second moment-like quantity:

$$v^2 = \sum_{k=0}^{\infty} k(k-1)p_k.$$

**Theorem 2.1** *Under mild conditions on the distribution $\{p_k\}$ and on the construction of the underlying random graph as described in Section 2.3, the basic reproduction number of the epidemic is:*

(i) **No testing.** $R_0 = \Big(\dfrac{\beta}{\beta+\gamma}\Big)\Big(\dfrac{(1-\mu)v^2}{\lambda}\Big)$.

(ii) **Random testing.** *If on average proportion $\rho$ of the population is tested daily:*

$$R_0(\rho) = \Big(\dfrac{\beta}{\beta+\gamma+\rho q(1-\delta)}\Big)\Big(\dfrac{(1-\mu)v^2}{\lambda}\Big).$$

The result of part (i) can be found, e.g., in [6]. The result of part (ii) follows from part (i) by the same reasoning that we used for the Bernoulli graph in Section 1.2.

The expressions for $R_0$ and $R_0(\theta)$ in equations (1) and (3) for the case of the simple Erdős-Rényi graph, are very similar to the corresponding expressions for $R_0$ and $R_0(\rho)$ in Theorem 2.1 for the case of the configuration model. Comparing the two it should be clear that, at least qualitatively, the effect of testing on the basic reproduction number should be very much the same in the more general case of networks with given degree distributions as it was in the case of the Erdős-Rényi graph. This is illustrated in the following section with an explicit example.

## 2.2 Example

Consider an epidemic on a random graph with degree distribution $\{p_k\}$ given by $p_0 = 0$ an,

$$p_k = B^{-1}\frac{1}{k^a}e^{-k/b}, \qquad k \geq 1,$$

with $B$ being the normalizing constant, and where for the parameters we take $a = 1.75$ and $b = 50$; see, e.g., [9] for motivation about this choice of $\{p_k\}$. This choice of $\{p_k\}$ is a heavy-tailed-like distribution $\propto 1/k^a$, with an exponential cut-off around $k = b$. It results in graphs with individuals having $\lambda \approx 3.5$ acquaintances on the average, but with a high variability.

We again assume that, initially, 1-in-10,000 people are infected ($\mu = 10^{-4}$), that the infectious period last for 7 days on the average ($\gamma = 1/7$), and that the contact rate $\beta$ varies between 0.0096 and 0.032. This means that an infected individual makes infectious contact, on the average, with $\lambda\beta$ others per day, where $\lambda\beta$ ranges between 0.112 and 0.336.

The same test-and-quarantine protocol as before is assumes, with a test having 70% sensitivity ($\delta = 0.3$), a testing rate of $\rho N$ per day with $\rho$ varying, and 75% quarantine compliance ($q = 0.75$).

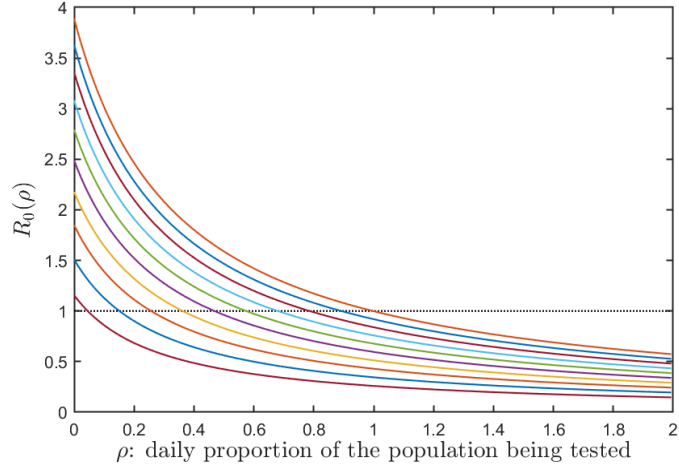

Figure 3: SIR epidemic on the configuration model: Effect of randomly testing a proportion $\rho$ of the population daily on the basic reproduction number $R_0$, in ten different cases, with contact rates $\beta = 0.0096, 0.0128, 0.0160, 0.0192, 0.0224, 0.0256, 0.0288, 0.0320, 0.0352, 0.0384$, and corresponding values for $R_0 = 1.1564, 1.5102, 1.8497, 2.1758, 2.4893, 2.7909, 3.0812, 3.3609, 3.6306, 3.8907$.

Figure 3 shows the resulting values of the basic reproduction number with and without testing. Once again we note that $\rho$, the average fraction of the population tested each day, can take arbitrary positive values, even above 1.

Finally we observe from Theorem 2.1 that, again, there is a critical testing rate $\rho^*$ such that, if $\rho > \rho^*$, the basic reproduction number $R_0(\rho)$ drops below 1:

$$\rho^* = \frac{1}{q(1-\delta)} \left[ \beta \Big( \frac{(1-\mu)v^2}{\lambda} - 1 \Big) - \gamma \right].$$

Figure 4 illustrates the relationship between $\rho^*$ and $R_0$. Obsrerve that the result is very similar to the corresponding plot in Figure 2 for the case of the Erdős-Rényi model, except that the rate of testing necessary to drop $R_0$ below 1 is slightly larger in the case of the confifuation model.

## 2.3 Mathematical assumptions

Here we give a precise description of the construction of the sequence of random graphs with degree distribution $\{p_k\}$, and we state the conditions required for the validity of the theorem.

For each $N$, we assume that there are, initially, $N_I$ infected individuals such that, for each $k \geq 0$, there $N_{I,k}$ individuals with degree $k$. Similarly, assume that there are $N_S = N - N_I$ susceptible individuals, with $N_{S,k}$ of them having degree $k$, for each $k \geq 0$. We assume that, as $N \to \infty$, $N_I/N \to \mu$ and $N_S/N \to (1-\mu)$, for some $\mu \in (0,1)$, and that,

$$\frac{N_{S,k}}{N_S} \to p_k, \qquad \frac{N_{I,k}}{N_I} \to p_k, \qquad k \geq 0,$$

$$\sum_{k=0}^{\infty} k \frac{N_{S,k}}{N_S} \to \sum_{k=0}^{\infty} k p_k = \lambda, \qquad \sum_{k=0}^{\infty} k \frac{N_{I,k}}{N_I} \to \sum_{k=0}^{\infty} k p_k = \lambda.$$

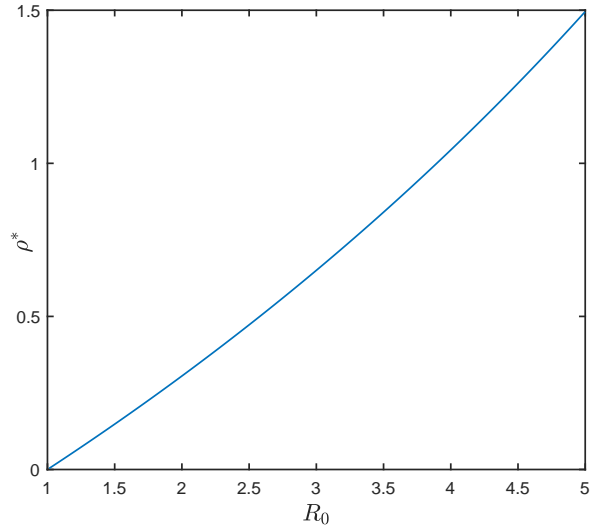

Figure 4: The critical testing rate $\rho^*$ for different scenarios with different values of $R_0$. The contact rate $\beta$ varies between 0.0083 and 0.0534, corresponding to values of $R_0$ between and 1 and 5, giving $\rho^*$ between 0 and 1.495.

This then implies that the average degree over all vertices also converges to $\lambda$. Writing $N_k$ for the total number of vertices with degree $k$,

$$\sum_{k=0}^{\infty} k \frac{N_k}{N} \to \lambda, \qquad \text{as } N \to \infty,$$

where we assume $\lambda \in (0, \infty)$. Also then we have,

$$\sum_{k=0}^{\infty} k \frac{N_{S,k}}{N} \to (1-\mu)\lambda, \qquad \sum_{k=0}^{\infty} k \frac{N_{I,k}}{N} \to \mu\lambda, \qquad \text{as } N \to \infty.$$

Two last technical assumptions that are required for the validity of Theorem 2.1, are that, as $N \to \infty$,

$$\max\{k \ ; \ N_{I,k} > 0\} = o(N), \qquad \text{and} \qquad \sum_{k=0}^{\infty} k^2 N_k = O(N).$$

Note that this last assumption implies that $\{p_k\}$ has a finite second moment.